\documentclass[conference,letterpaper,10pt]{IEEEtran}
\IEEEoverridecommandlockouts

\usepackage{cite}
\usepackage{amsmath,amssymb,amsfonts}
\usepackage{algorithmic}
\usepackage{graphicx}
\usepackage{textcomp}
\usepackage{xcolor}
\usepackage{booktabs}
\usepackage{multirow}
\def\BibTeX{{\rm B\kern-.05em{\sc i\kern-.025em b}\kern-.08em
    T\kern-.1667em\lower.7ex\hbox{E}\kern-.125emX}}
\begin{document}
\bstctlcite{IEEEexample:BSTcontrol}

\title{Spatially Distributed Task-Oriented Compression for Multi-Emitter Localization and Characterization with Spectral Overlap
}

\author{
\IEEEauthorblockN{H. Nazim Bicer and J. Nicholas Laneman}
\IEEEauthorblockA{
\textit{Dept. of Electrical Engineering}\\
\textit{University of Notre Dame}\\
Notre Dame, IN, USA\\
\{hbicer, jnl\}@nd.edu
}
}

\maketitle

\begin{abstract}
Radio frequency spectrum awareness requires the ability to detect, localize, and characterize emitters in dense and contested wireless environments.
In this work, we propose a task-oriented distributed compression framework for joint multi-emitter localization and characterization using spatially distributed receivers. Each receiver observes a short window of complex IQ samples, converts the observation to a time--frequency representation, and encodes it into a compact latent vector. A central fusion decoder combines the receiver latents to estimate an unordered set of active emitters, including their locations, center-frequency offsets, occupied bandwidths, and waveform families. A permutation-invariant training objective is used to handle the arbitrary ordering of emitters and predictions. Experiments on synthetic multi-emitter scenes with spectral overlap show that even extremely compact receiver-side representations can preserve useful information for emitter counting and waveform-family estimation. However, accurate localization and spectral-parameter regression require larger latent dimensions. Increasing the receiver latent dimension from $d_{\mathrm{rx}}=1$ to $d_{\mathrm{rx}}=16$ provides the largest improvement, while further increasing to $d_{\mathrm{rx}}=64$ gives smaller gains. These results demonstrate the potential of learned task-oriented compression for communication-efficient distributed spectrum awareness.
\end{abstract}


\section{Introduction}

Radio frequency (RF) spectrum awareness is becoming increasingly critical as wireless deployments grow denser and spectrum access becomes more contested.
In many commercial systems, interference is managed through scheduling, spectrum planning, coexistence rules, or prior coordination among authorized users \cite{Hamza_etal,Boudreau_etal,Akyildiz_etal}.
These mechanisms are effective when transmitters are known and cooperative.
However, unauthorized, uncoordinated, malfunctioning, or adversarial emitters can still disrupt systems, including critical infrastructure, public safety networks, airports, and other operational environments \cite{Pirayesh_etal,Liu_etal}.
Rapidly detecting, localizing, and characterizing such emitters is therefore an important capability for spectrum security and resilient wireless operation.

A single high-quality receiver can often detect the presence of an interferer when the received signal-to-noise ratio is sufficient \cite{SNR_walls,Yucek_2009}.
However, single-receiver observations provide limited spatial information, especially when multiple emitters are active in a complex propagation environment.
Distributed sensing offers a natural way to address this limitation.
By observing the same RF scene from multiple spatially separated locations, a receiver network can exploit spatial diversity and recover information that is not available from any individual receiver alone \cite{Ghasemi_2005,Akyildiz2011}.
At the same time, low-cost software-defined radios (SDRs) and embedded processors increasingly support local processing of wideband observations, making cooperative spectrum sensing practical at the network edge.

Modern SDRs can collect complex in-phase and quadrature (IQ) samples, which preserve spectral, temporal, and modulation-dependent structure beyond received power alone.
This enables spectrum awareness tasks that go beyond detection or localization, including estimation of the number of active emitters, their locations, carrier frequencies, bandwidths, and waveform families.
However, directly forwarding raw IQ samples from many receivers to a central processor can require substantial communication bandwidth, particularly for wideband sensing.
This motivates task-oriented distributed compression, where each receiver transmits a compact representation designed specifically for the downstream inference task rather than a faithful reconstruction of the raw observation \cite{ShaoEdgeInf,ShaoTrans}.

In this paper, we study communication-efficient distributed spectrum awareness using multiple spatially separated SDR-like receivers.
Each receiver observes a short window of IQ samples and maps its local observation to a low-dimensional latent representation using a neural encoder.
A central neural decoder then fuses the receiver latents and estimates the active emitter set.
Unlike formulations that address detection, localization, or waveform classification separately, we consider a joint multi-emitter setting in which the number of emitters, spatial locations, carrier frequencies, bandwidths, and waveform families are all unknown.
The output is treated as an unordered set, allowing the model to estimate multiple active emitters without imposing an artificial ordering on the scene \cite{DeepSets}.
During training, we handle the resulting assignment ambiguity with a permutation-invariant objective that minimizes the loss over possible matches between predicted and true emitters, following the same principle as permutation-invariant training for multi-source separation \cite{Yu_PIT}.

The main contributions of this work are as follows.
First, we formulate distributed spectrum awareness as a joint emitter-set estimation problem from compressed multi-receiver IQ observations.
Second, we propose a task-oriented neural compression architecture in which receiver-side encoders produce compact latents that are fused by a central neural decoder for joint emitter localization and characterization.
Third, we use a permutation-invariant training objective to handle the unordered nature of multi-emitter outputs, including scenes with partially or strongly overlapping spectra.
Finally, we evaluate how receiver-side latent dimensionality and maximum emitter count affect cardinality estimation, localization accuracy, and spectral-parameter recovery.

\section{Related Work}
\label{sec:related_work}

The communication-efficient distributed inference problem considered in this paper is closely related to classical multiterminal source coding.
In the CEO problem \cite{CEO}, multiple agents observe noisy versions of an underlying source and send rate-limited descriptions to a central estimator.
The distributed information bottleneck framework extends this idea by compressing multiple observations into representations that collectively preserve information relevant to a target variable \cite{EstellaAguerri2018DIB}.
This perspective naturally fits distributed spectrum awareness: each receiver observes a different view of the same RF scene, while the fusion center only requires information relevant to estimating the active emitter set rather than reconstructing the raw observations.

Recent task-oriented communication methods have translated these ideas into neural edge-inference architectures.
The work in \cite{ShaoEdgeInf} studied information-bottleneck-based feature encoding for single-device edge inference, where compact task-relevant features are transmitted instead of full observations.
A subsequent study extended this setting to cooperative multi-device inference using distributed feature encoding to reduce communication overhead \cite{ShaoTrans}.
These works motivate communication-efficient fusion, but focus mainly on image-based inference rather than RF spectrum awareness.

Deep-learning-based multiple-transmitter localization has also been studied for spectrum monitoring.
DeepTxFinder \cite{DeepTXFinder} uses distributed received-power measurements and a convolutional neural network to estimate the number and locations of unknown transmitters.
DeepMTL \cite{DeepMTL} and DeepMTL Pro \cite{DeepMTLPro} formulate transmitter localization as a learned spatial inference problem from received-power measurements, with DeepMTL Pro additionally estimating transmitter power.
These methods demonstrate the value of distributed sensing and deep learning, but they primarily rely on power-based measurements and therefore do not exploit the richer spectral, temporal, and waveform-dependent information available in wideband complex IQ observations.

Complementary work on edge-based RF signal detection and characterization developed an RFSoC-deployable neural model that operates on Fourier domain passband RF data to detect and classify RF signals in real time while estimating center frequency and bandwidth \cite{parpart2025passband}.
This demonstrates the feasibility of neural spectrum awareness processing on deployable hardware, but it considers an individual edge platform rather than cooperative sensing across spatially distributed receivers and does not address emitter localization or distributed task-oriented compression.

Our work lies at the intersection of these directions.
Motivated by distributed information bottleneck and task-oriented communication principles, we compress observations locally at multiple receivers and fuse the resulting latent representations centrally.
Relative to power-based localization methods, we use IQ-derived observations to jointly estimate emitter cardinality, locations, center-frequency offsets, bandwidths, and waveform families.
Relative to single-platform edge RF detection methods, we study a spatially distributed latent-bottleneck architecture for unordered multi-emitter scenes with potentially overlapping spectra.

\section{Problem Formulation}
\label{sec:problem_formulation}

We consider a two-dimensional monitoring region
\(\mathcal{A} \subset \mathbb{R}^{2}\) containing \(M\) spatially
distributed receivers. Receiver \(i \in \{1,\ldots,M\}\) is located at
a known position \(\mathbf{r}_i \in \mathcal{A}\).
The receiver locations are fixed across all scenes, and receiver observations are provided to the fusion center in a consistent order.

All receivers observe the same RF scene over a common complex-baseband observation window during an interval \(T_{\mathrm{obs}}\).
With sampling rate \(f_s\), receiver \(i\) obtains
\(N=f_sT_{\mathrm{obs}}\) complex samples, where \(N\) is assumed to be
an integer:
\begin{equation}
    \mathbf{x}_i
    =
    \left[
    x_i[0],x_i[1],\ldots,x_i[N-1]
    \right]^{\mathsf T}
    \in \mathbb{C}^{N}.
    \label{eq:rx_iq_vector}
\end{equation}
The modeled complex-baseband observation window is
\begin{equation}
    \mathcal{F}
    =
    \left[
    -\frac{f_s}{2},
    \frac{f_s}{2}
    \right).
    \label{eq:monitored_band}
\end{equation}
Thus, frequencies in this work represent offsets within the sampled
baseband window rather than absolute RF carrier frequencies.

Each observed scene contains an unknown number \(K\) of simultaneously
active emitters, with \(1 \leq K \leq K_{\max}\). The physical
parameters governing the signal generated by emitter \(k\) are
represented by
\begin{equation}
    \boldsymbol{\xi}_k
    =
    \left(
    \mathbf{p}_k,\,
    \nu_k,\,
    B_k,\,
    P_k,\,
    g_k,\,
    q_k
    \right),
    \label{eq:physical_emitter_parameters}
\end{equation}
where \(\mathbf{p}_k \in \mathcal{A}\) is the emitter location,
\(\nu_k \in \mathcal{F}\) is its center-frequency offset within the
sampled baseband window, \(B_k\) is its occupied bandwidth, \(P_k\) is
its transmit-power parameter, \(g_k \in \mathcal{G}\) is its
waveform-family label, and \(q_k \in \mathcal{Q}\) is its modulation
label. The spectral placement of each emitter satisfies
\(\nu_k \in [-f_s/2+B_k/2,\, f_s/2-B_k/2)\), so that its occupied
bandwidth remains inside the observation window. 
Different emitters may have partially or fully overlapping occupied frequency intervals, may use the same waveform family or modulation type,
and may transmit with different powers.

In this work, transmit power and modulation type are treated as nuisance
variables: they influence the receiver observations but are not included
in the target output. The task-relevant parameter vector associated with
emitter \(k\) is therefore
\begin{equation}
    \boldsymbol{\theta}_k
    =
    \left(
    \mathbf{p}_k,\,
    \nu_k,\,
    B_k,\,
    g_k
    \right),
    \label{eq:target_emitter_parameters}
\end{equation}
and the desired scene-level output is the unordered emitter set
\begin{equation}
    \Theta
    =
    \left\{
    \boldsymbol{\theta}_k
    \right\}_{k=1}^{K}.
    \label{eq:target_emitter_set}
\end{equation}

Let \(s_{g_k,q_k}[n;B_k]\) denote a unit-power complex-baseband
waveform whose structure depends on the waveform family, modulation
type, and occupied bandwidth of emitter \(k\). Including its
center-frequency offset within the sampled baseband window, define
\begin{equation}
    u_k[n]
    =
    s_{g_k,q_k}[n;B_k]\,
    e^{j2\pi \nu_k n/f_s},
    \qquad
    n=0,\ldots,N-1.
    \label{eq:emitter_signal}
\end{equation}

Let \(d_{i,k}=\|\mathbf{p}_k-\mathbf{r}_i\|_2\) denote the Euclidean
distance between emitter \(k\) and receiver \(i\). The simulated channel
is designed to isolate localization information arising from
receiver-dependent received powers. Specifically, the received IQ sample
at receiver \(i\) is modeled as
\begin{equation}
    x_i[n]
    =
    \sum_{k=1}^{K}
    \sqrt{P_k}\,
    \alpha_{i,k}\,
    u_k[n]
    +
    w_i[n],
    \qquad
    n=0,\ldots,N-1.
    \label{eq:receiver_observation_model}
\end{equation}
where \(w_i[n] \sim \mathcal{CN}(0,\sigma_w^2)\) denotes receiver
thermal noise modeled as complex AWGN, and \(\alpha_{i,k}\) is the
distance-dependent amplitude attenuation for the link between emitter
\(k\) and receiver \(i\).

The amplitude attenuation is modeled using a reference-distance
power-law relationship \cite[Sec.~2.5]{GoldsmithWireless}:
\begin{equation}
    \alpha_{i,k}
    =
    \left(
    \frac{\max(d_{i,k},d_0)}{d_0}
    \right)^{-\eta/2},
    \label{eq:channel_amplitude}
\end{equation}
where \(d_0\) is the reference distance and \(\eta\) is the path-loss
exponent. Accordingly, the received power contribution of emitter
\(k\) at receiver \(i\) is
\(P^{\mathrm{rx}}_{i,k}=P_k\alpha_{i,k}^{2}\), with all power quantities
understood in linear units.

No propagation delay, phase rotation, multipath effect, shadowing, or
receiver synchronization impairment is applied in the present simulation
model. Consequently, emitter localization is driven only by the
receiver-dependent power variations induced by the known receiver
geometry. The complex time--frequency observations retain spectral and
waveform-dependent structure for estimating center-frequency offset,
occupied bandwidth, and waveform family, but they do not contain
additional modeled phase- or time-difference-based localization cues.

The objective is to infer \(\Theta\) from the distributed observations
\(\{\mathbf{x}_i\}_{i=1}^{M}\) while limiting the dimensionality of the
information provided by each receiver to the fusion center. Specifically,
receiver \(i\) forms a latent representation
\begin{equation}
    \mathbf{z}_i
    =
    \operatorname{Enc}_{\phi_i}
    \left(
    \mathcal{T}
    \left(
    \mathbf{x}_i
    \right)
    \right)
    \in \mathbb{R}^{d_{\mathrm{rx}}},
    \label{eq:abstract_encoder}
\end{equation}
where \(\phi_i\) denotes the parameters of receiver \(i\)'s encoder and
\(\mathcal{T}(\cdot)\) represents the local observation transform. The
specific transform and neural encoder realization are defined in
Section~\ref{sec:proposed_method}. The receivers employ the same encoder
architecture but do not share encoder parameters, allowing
receiver-specific representations under the fixed receiver geometry
considered in this work.

The fusion center estimates the emitter set according to
\begin{equation}
    \widehat{\Theta}
    =
    \operatorname{Dec}_{\psi}
    \left(
    \mathbf{z}_1,\ldots,\mathbf{z}_M
    \right),
    \label{eq:abstract_decoder}
\end{equation}
where \(\psi\) denotes the fusion-decoder parameters.
Since the latent vectors are provided to the decoder in a consistent
receiver order, the decoder can learn receiver-dependent power
relationships associated with the fixed sensing geometry without taking
receiver coordinates as additional numerical inputs.

Here, \(d_{\mathrm{rx}}\) controls the dimensionality of the
receiver-side latent bottleneck. Since the latent vectors are not
quantized or entropy coded, \(d_{\mathrm{rx}}\) characterizes
representation compactness rather than an explicit communication bit rate.

The training objective must be invariant to permutations of emitter indices. We therefore minimize an expected set distortion that accounts for existence, waveform-family classification, and continuous parameter-estimation errors:
\begin{equation}
    \underset{\{\phi_i\}_{i=1}^{M},\,\psi}
    {\operatorname{minimize}}
    \quad
    \mathbb{E}
    \left[
    \mathcal{D}_{\mathrm{set}}
    \left(
    \Theta,\widehat{\Theta}
    \right)
    \right].
    \label{eq:problem_objective}
\end{equation}

\section{Proposed Method}
\label{sec:proposed_method}

To address the distributed estimation problem in Section~\ref{sec:problem_formulation}, we use a neural architecture with receiver-specific local encoders and a central fusion decoder.
Each receiver maps its local IQ observation to a compact latent vector.
The fusion decoder then combines the ordered collection of receiver latents to estimate the unordered set of active emitters.

\subsection{Local Time--Frequency Encoding}
\label{subsec:local_encoding}

In Section~\ref{sec:problem_formulation}, the receiver-side encoder was defined using an abstract local transform $\mathcal{T}(\cdot)$.
In this work, we set $\mathcal{T}(\cdot)$ to a time--frequency representation based on the short-time Fourier transform (STFT).
For receiver $i$, the complex IQ sequence $\mathbf{x}_i$ is mapped to
\begin{equation}
    \mathbf{S}_i
    =
    \operatorname{STFT}
    \left(
    \mathbf{x}_i
    \right)
    \in
    \mathbb{C}^{N_f \times N_t},
    \label{eq:local_stft}
\end{equation}
where $N_f$ and $N_t$ denote the number of frequency bins and time frames, respectively.
The real and imaginary components are then stacked as two input channels,
\begin{equation}
    \mathbf{X}_i
    =
    \mathcal{T}
    \left(
    \mathbf{x}_i
    \right)
    =
    \left[
    \operatorname{Re}\{\mathbf{S}_i\};
    \operatorname{Im}\{\mathbf{S}_i\}
    \right]
    \in
    \mathbb{R}^{2 \times N_f \times N_t}.
    \label{eq:local_stft_tensor}
\end{equation}
This representation preserves spectral placement and complex-valued time--frequency structure relevant to estimating emitter bandwidth, center-frequency offset, and waveform family.
Time--frequency representations have also been shown to outperform raw IQ inputs in a related neural RF classification task \cite{GlugeDroneRF}.

Each receiver applies a local encoder to produce a $d_{\mathrm{rx}}$-dimensional latent vector,
\begin{equation}
    \mathbf{z}_i
    =
    \operatorname{Enc}_{\phi_i}
    \left(
    \mathbf{X}_i
    \right)
    \in
    \mathbb{R}^{d_{\mathrm{rx}}},
    \qquad
    i=1,\ldots,M.
    \label{eq:local_latent}
\end{equation}
All receivers use the same encoder architecture, but the encoder parameters are not shared.
This allows each receiver to learn a receiver-specific representation under the fixed sensing geometry.
The encoder is implemented as a residual convolutional network with asymmetric downsampling.
The time dimension is reduced more aggressively than the frequency dimension before projection to the latent vector.

\subsection{Central Fusion and Set Prediction}
\label{subsec:central_fusion}

The fusion center concatenates the receiver latents in the fixed receiver order,
\begin{equation}
    \mathbf{z}
    =
    \left[
    \mathbf{z}_1^{\mathsf T},
    \ldots,
    \mathbf{z}_M^{\mathsf T}
    \right]^{\mathsf T}
    \in
    \mathbb{R}^{M d_{\mathrm{rx}}}.
    \label{eq:fused_latent}
\end{equation}
A multilayer perceptron maps the fused latent vector to $K_{\max}$ candidate emitter slots,
\begin{equation}
    \widehat{\mathbf{Y}}
    =
    \operatorname{Dec}_{\psi}
    \left(
    \mathbf{z}
    \right)
    \in
    \mathbb{R}^{K_{\max} \times 6}.
    \label{eq:decoder_prediction}
\end{equation}
Each row of $\widehat{\mathbf{Y}}$ corresponds to one candidate emitter slot.
For slot $j$, the decoder output is
\begin{equation}
    \widehat{\mathbf{y}}_j
    =
    \left[
    \widehat{e}_j,\,
    \widehat{p}_{x,j},\,
    \widehat{p}_{y,j},\,
    \widehat{\nu}_j,\,
    \widehat{B}_j,\,
    \widehat{c}_j
    \right],
    \label{eq:predicted_slot}
\end{equation}
where $\widehat{e}_j$ is an existence logit.
The pair $(\widehat{p}_{x,j},\widehat{p}_{y,j})$ is the estimated emitter location.
The quantity $\widehat{\nu}_j$ is the estimated center-frequency offset, and $\widehat{B}_j$ is the estimated occupied bandwidth.
The quantity $\widehat{c}_j$ is a waveform-family logit.
In the experiments considered here, the waveform-family label is binary, so a single logit is sufficient.
The continuous quantities are normalized during training and converted back to physical units for evaluation.

\subsection{Permutation-Invariant Training}
\label{subsec:permutation_training}

The decoder always produces $K_{\max}$ slots, while a scene contains only $K \leq K_{\max}$ active emitters.
We therefore represent each ground-truth scene using padded target slots
\begin{equation}
    \begin{aligned}
    \mathbf{y}_k
    &=
    \left[
    e_k,\,
    p_{x,k},\,
    p_{y,k},\,
    \nu_k,\,
    B_k,\,
    c_k
    \right],
    \\
    &\hspace{2em} k=1,\ldots,K_{\max}.
    \end{aligned}
    \label{eq:target_slot}
\end{equation}
where $e_k=1$ for active emitters and $e_k=0$ for padded inactive slots.
For inactive slots, the remaining target entries are set to zero by convention.
These zero-padded entries are not treated as physical emitter parameters, since the corresponding waveform-family and regression losses are masked by $e_k$.
The label $c_k \in \{0,1\}$ denotes the binary encoding of the waveform-family label $g_k$ when $e_k=1$.

Since emitter ordering is arbitrary, predicted and target slots are matched using a permutation-invariant cost.
The cost of assigning predicted slot $j$ to target slot $k$ is
\begin{equation}
    \begin{aligned}[t]
    C(k,j)
    =
    &\lambda_{\mathrm{e}}
    \ell_{\mathrm{BCE}}
    \left(
    \widehat{e}_j,e_k
    \right)
    +
    e_k\lambda_{\mathrm{fam}}
    \ell_{\mathrm{BCE}}
    \left(
    \widehat{c}_j,c_k
    \right)
    \\
    &+
    e_k
    \Big[
    \lambda_x
    \ell_{\mathrm{reg}}
    \left(
    \widehat{p}_{x,j},p_{x,k}
    \right)
    +
    \lambda_y
    \ell_{\mathrm{reg}}
    \left(
    \widehat{p}_{y,j},p_{y,k}
    \right)
    \\
    &\qquad+
    \lambda_{\mathrm{fc}}
    \ell_{\mathrm{reg}}
    \left(
    \widehat{\nu}_j,\nu_k
    \right)
    +
    \lambda_{\mathrm{bw}}
    \ell_{\mathrm{reg}}
    \left(
    \widehat{B}_j,B_k
    \right)
    \Big].
    \end{aligned}
    \label{eq:pairwise_matching_cost}
\end{equation}
where $\ell_{\mathrm{BCE}}$ is binary cross-entropy applied to logits and
$\ell_{\mathrm{reg}}$ is the Smooth $L_1$ loss, i.e., a Huber-type regression loss applied to normalized continuous targets.
The factor $e_k$ ensures that waveform-family and continuous-parameter errors are only penalized for active emitters.

The optimal assignment is
\begin{equation}
    \pi^{\star}
    =
    \underset{\pi \in \mathfrak{S}_{K_{\max}}}
    {\operatorname{argmin}}
    \;
    \sum_{k=1}^{K_{\max}}
    C\left(k,\pi(k)\right),
    \label{eq:optimal_permutation}
\end{equation}
where $\mathfrak{S}_{K_{\max}}$ is the set of all permutations of the $K_{\max}$ output slots.
In the experiments, we consider $K_{\max}\in\{3,4,5\}$.
Therefore, the exact minimum is obtained by evaluating $6$, $24$, and $120$ assignments, respectively.

Using the assignment $\pi^\star$, the training loss is
\begin{align}
    \mathcal{L}
    =
    &\lambda_{\mathrm{e}}\mathcal{L}_{\mathrm{e}}
    +
    \lambda_{\mathrm{fam}}\mathcal{L}_{\mathrm{fam}}
    +
    \lambda_x\mathcal{L}_x
    +
    \lambda_y\mathcal{L}_y
    \nonumber\\
    &+
    \lambda_{\mathrm{fc}}\mathcal{L}_{\mathrm{fc}}
    +
    \lambda_{\mathrm{bw}}\mathcal{L}_{\mathrm{bw}},
    \label{eq:training_loss}
\end{align}
where $\mathcal{L}_{\mathrm{e}}$ is computed over all matched slots.
The remaining loss terms are computed using the matched target existence indicators.
Specifically, the waveform-family and continuous-parameter losses are masked by $e_k$ and averaged only over active target slots.
Thus, inactive padded slots contribute only to the existence loss.
The latent dimension $d_{\mathrm{rx}}$ controls the receiver-side bottleneck, and $K_{\max}$ controls the maximum scene complexity considered during training and evaluation.

\begin{figure*}[t]
    \centering
    \includegraphics[width=0.85\textwidth]{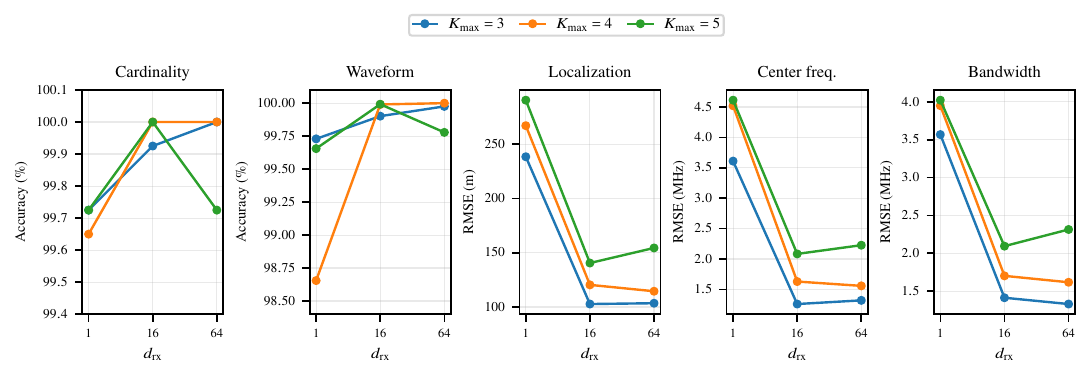}
\caption{Test set performance versus receiver latent dimension $d_{\mathrm{rx}}$ for $K_{\max}\in\{3,4,5\}$.
The panels show cardinality accuracy, waveform-family accuracy, localization RMSE, center-frequency RMSE, and bandwidth RMSE.}
    \label{fig:summary_results}
\end{figure*}

\section{Experiments and Results}
\label{sec:experiments_results}

\subsection{Experimental Setup}
\label{subsec:experimental_setup}

We evaluate the proposed architecture using synthetic multi-emitter RF scenes generated according to the model in Section~\ref{sec:problem_formulation}.
The monitored region is an $800\,\mathrm{m}\times800\,\mathrm{m}$ area with $M=4$ fixed receivers located at $(50,100)$, $(125,700)$, $(700,150)$, and $(675,600)$ m.
Each receiver observes $T_{\mathrm{obs}}=1$ ms of complex baseband IQ samples at $f_s=40$ MHz, corresponding to a modeled baseband window of $[-20,20)$ MHz.
Each scene contains between one and $K_{\max}$ active emitters.
Emitters are assigned a center-frequency offset, occupied bandwidth, waveform family, modulation type, and transmit-power parameter.
Bandwidths are sampled between $1$ MHz and $16$ MHz, and center frequencies are chosen so that the occupied band remains within the observation window.
The waveform-family label is binary, corresponding to OFDM or single-carrier signals, while modulation type and transmit power are nuisance variables.

Separate datasets are generated for $K_{\max}\in\{3,4,5\}$.
For each value of $K_{\max}$, we use $40{,}000$ training scenes, $4{,}000$ validation scenes, and $4{,}000$ test scenes.
The propagation model uses distance-dependent attenuation with path-loss exponent $\eta=2.7$ and excludes delay, multipath, shadowing, and phase-based localization cues.
Thus, localization information is limited to receiver-dependent power variation under the fixed receiver geometry.

Each receiver IQ sequence is converted to a complex STFT using a Hann window with $N_f=256$ frequency bins and hop length $160$, yielding $N_t=249$ time frames.
The real and imaginary STFT components are stacked as two channels, producing a $2\times256\times249$ tensor per receiver.
We train models with receiver latent dimensions $d_{\mathrm{rx}}\in\{1,16,64\}$, corresponding to fused latent dimensions of $4$, $64$, and $256$ at the central decoder.

Each receiver uses an independent residual convolutional encoder with the same architecture but separately learned parameters. The encoder applies asymmetric downsampling, adaptive average pooling, and a fully connected projection to produce $\mathbf{z}_i\in\mathbb{R}^{d_{\mathrm{rx}}}$.
The fusion center concatenates the four receiver latents in fixed receiver order and applies an MLP with hidden dimensions $(256,256,128)$.
The decoder outputs $K_{\max}$ candidate emitter slots, each containing an existence logit, two normalized location coordinates, a normalized center-frequency offset, a normalized bandwidth, and a waveform-family logit.

For each pair of $K_{\max}$ and $d_{\mathrm{rx}}$, a separate model is trained using the permutation-invariant loss in Section~\ref{subsec:permutation_training}.
Exact assignment evaluates all $K_{\max}!$ slot permutations during training.
All models are trained for $200$ epochs using AdamW with learning rate $10^{-3}$ and batch size $64$.
The checkpoint with the lowest validation loss is used for test set evaluation.

\subsection{Results and Discussion}
\label{subsec:results_discussion}

We evaluate scene-level cardinality accuracy, waveform-family accuracy, localization RMSE, center-frequency RMSE, and bandwidth RMSE. Waveform-family accuracy and continuous-parameter errors are computed after permutation-invariant matching between predicted and true active emitters.

Figure~\ref{fig:summary_results} shows that the receiver-side bottleneck strongly affects continuous estimation.
With $d_{\mathrm{rx}}=1$, the fused representation is only four-dimensional, and the model loses substantial fine-grained information.
Increasing the latent dimension to $d_{\mathrm{rx}}=16$ gives the largest improvement across all values of $K_{\max}$.
Localization RMSE decreases from $238.4$, $267.0$, and $290.5$ m for $K_{\max}=3,4,5$, respectively, at $d_{\mathrm{rx}}=1$ to $102.7$, $120.3$, and $140.5$ m at $d_{\mathrm{rx}}=16$, while center-frequency and bandwidth errors decrease from approximately $3.5$--$4.6$ MHz to about $1$--$2$ MHz.
Increasing the latent dimension further to $d_{\mathrm{rx}}=64$ gives smaller and less consistent gains, suggesting that $d_{\mathrm{rx}}=16$ captures most of the useful information while keeping the receiver representation compact.

Performance also degrades as $K_{\max}$ increases.
This is expected because larger values of $K_{\max}$ allow more simultaneous emitters, increasing spatial ambiguity, spectral overlap, and assignment difficulty.
The effect is most visible in localization and bandwidth RMSE, where the $K_{\max}=5$ case is consistently more challenging than the $K_{\max}=3$ case. 

\begin{figure*}[t]
    \centering
    \includegraphics[width=\textwidth]{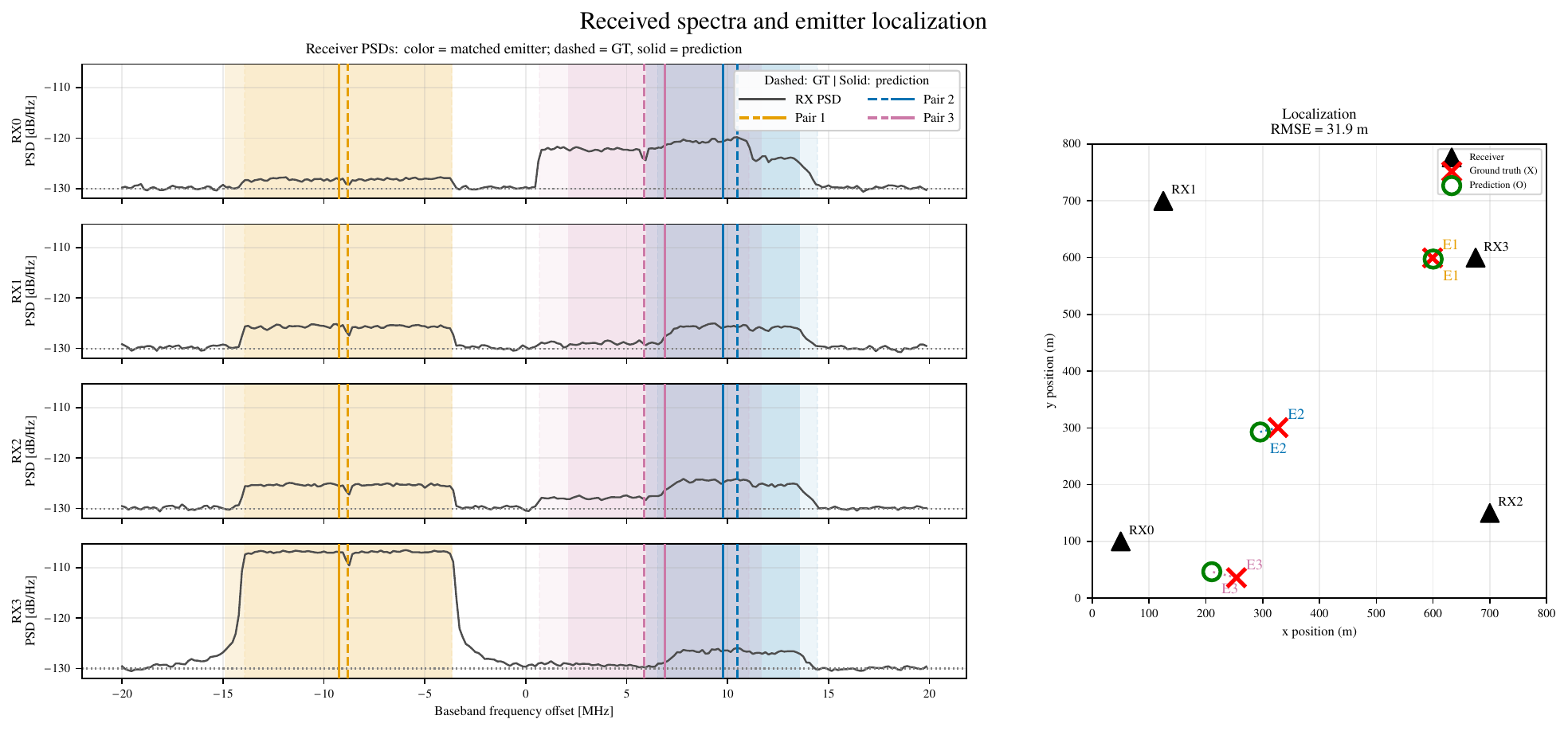}
\caption{Representative test example with three active emitters using $d_{\mathrm{rx}}=64$. Receiver PSDs show ground-truth bandwidths and ground-truth/predicted center frequencies; the map shows receiver, true-emitter, and predicted-emitter locations. Two positive-frequency emitters partially overlap in spectrum. The active-emitter RMSE values are $31.9$ m for localization, $0.761$ MHz for center frequency, and $0.753$ MHz for bandwidth.}

    \label{fig:sample_inference}
\end{figure*}

Figure~\ref{fig:sample_inference} shows a representative inference example with three active emitters, including two emitters with partially overlapping spectra.

The proposed method has several limitations. The trained model is tied to the fixed receiver geometry and simulated propagation statistics, so changes in receiver placement or RF environment may require retraining or fine-tuning. The decoder also assumes a known $K_{\max}$, fixed receiver ordering, and no receiver-coordinate inputs. Finally, $d_{\mathrm{rx}}$ measures latent compactness rather than explicit communication rate, since quantization, entropy coding, channel effects, and hardware latency are not modeled.

\section{Conclusion}
\label{sec:conclusion}

We proposed a task-oriented distributed compression framework for joint localization and characterization of multiple RF emitters from spatially distributed IQ observations. Each receiver encodes a local time--frequency representation into a compact latent vector, and a central decoder fuses the receiver latents to estimate an unordered emitter set. Experiments on synthetic multi-emitter scenes show that receiver-side latent dimensionality strongly affects continuous parameter estimation: increasing $d_{\mathrm{rx}}$ from $1$ to $16$ provides the largest performance gain, while further increasing to $64$ yields smaller improvements. These results suggest that compact learned receiver representations can preserve useful task-relevant information for distributed spectrum awareness. Future work will incorporate explicit bit-level rate constraints, quantization, geometry-aware fusion, and more realistic propagation and hardware effects.

\bibliographystyle{IEEEtran}
\bibliography{refs}

\end{document}